\documentclass[conference]{IEEEtran}
\IEEEoverridecommandlockouts
\usepackage{cite}
\usepackage{amsmath,amssymb,amsfonts}
\usepackage{algorithmic}
\usepackage{graphicx}
\usepackage{textcomp}
\usepackage[bookmarks=false]{hyperref}
\usepackage{xcolor}
\usepackage{siunitx}
\usepackage[ancient]{flushend}
\def\BibTeX{{\rm B\kern-.05em{\sc i\kern-.025em b}\kern-.08em
    T\kern-.1667em\lower.7ex\hbox{E}\kern-.125emX}}
\begin{document}
\title{Machine Learning-Based Unbalance Detection of a Rotating Shaft Using Vibration Data}

\author{
\IEEEauthorblockN{Oliver Mey, Willi Neudeck, Andr\'{e} Schneider and Olaf Enge-Rosenblatt}
\IEEEauthorblockA{\textit{Fraunhofer IIS/EAS, Fraunhofer Institute for Integrated Circuits} \\
\textit{Division Engineering of Adaptive Systems}\\
Dresden, Germany \\
oliver.mey@eas.iis.fraunhofer.de}
}

\IEEEoverridecommandlockouts
\IEEEpubid{\makebox[\columnwidth]{\copyright2020
IEEE \hfill} \hspace{\columnsep}\makebox[\columnwidth]{ }} 
\maketitle
\begin{abstract}
Fault detection at rotating machinery with the help of vibration sensors offers the possibility to detect damage to machines at an early stage and to prevent production downtimes by taking appropriate measures. The analysis of the vibration data using methods of machine learning promises a significant reduction in the associated analysis effort and a further improvement in diagnostic accuracy. Here we publish a dataset which is used as a basis for the development and evaluation of algorithms for unbalance detection. For this purpose, unbalances of various sizes were attached to a rotating shaft using a 3D-printed holder. In a speed range from approx. 630 RPM to 2330 RPM, three sensors were used to record vibrations on the rotating shaft at a sampling rate of 4096 values per second. A development and an evaluation dataset are available for each unbalance strength. Using the dataset recorded in this way, fully connected and convolutional neural networks, Hidden Markov Models and Random Forest classifications on the basis of automatically extracted time series features were tested. With a prediction accuracy of \SI{98.6}{\percent} on the evaluation dataset, the best result could be achieved with a fully-connected neural network that receives the scaled FFT-transformed vibration data as input.
\end{abstract}

\section{Introduction}
Progress in the field of machine learning has led to impressive results in recent years, for example in the areas of image recognition\cite{lecun1998,lecun2015, krizhvsky2017, Szegedy2017}, natural language processing \cite{bojanowski2017,conneau2017,devlin2018,young2018,yang2019} or reinforcement learning \cite{mnih2013,mnih2015,silver2017,silver2018}. In addition to these examples, which are very present in the media, these algorithms also offer great potential for industrial applications \cite{li2017,zhao2019,bayer2013,lessmeier2014,enge2016}. For example, the analysis of vibrations on rotating shafts to detect unbalances or to detect damage to roller bearings has proven to be very promising \cite{dadouche2008,gopinath2010,janssens2016,carbajal2016,tahir2016,chen2017,liu2018,guo2019,tian2019}. Here, we focus on the first mentioned use case. Unbalances on rotating shafts can cause decreased lifetimes of bearings or other parts of the machinery and, therefore, lead to additional costs. Hence, early detection of unbalances helps to minimize maintenance expenses, to avoid unnecessary production stops and to increase the service life of machines. Algorithmic detection of unbalances is accompanied with the least additional effort. The automation achieved in this way also enables live analysis of streamed data, which means that unbalances can be detected and corrected with almost no time delay, even before potential damage to the drive train occurs.

We observe that there are only a few publicly available condition monitoring (CM) datasets with the help of which algorithms can be tested and compared. There are e.g. datasets for CM with hydraulic systems \cite{helwig2015} and for detecting bearing damage \cite{lessmeier2016,cwru_bdc,huang2018}, but there seems to be no dataset for detecting unbalances, which in turn can be a cause of bearing damage. For this reason, we publish a dataset for the detection of unbalances based on vibration data along with this study (available in the Fraunhofer Fordatis database \cite{dataset}). In addition, we carry out analyses to determine which algorithms can detect the unbalance as accurately as possible and up to which unbalance strength these can still be reliably recognized by each algorithm. The Python code used for the investigations conducted in this study is open-sourced in a Github repository \cite{our_github}.
\section{Measurement Setup}
\label{measurement_section}
\begin{figure}[htbp]
\centerline{\includegraphics[width=1.0\linewidth]{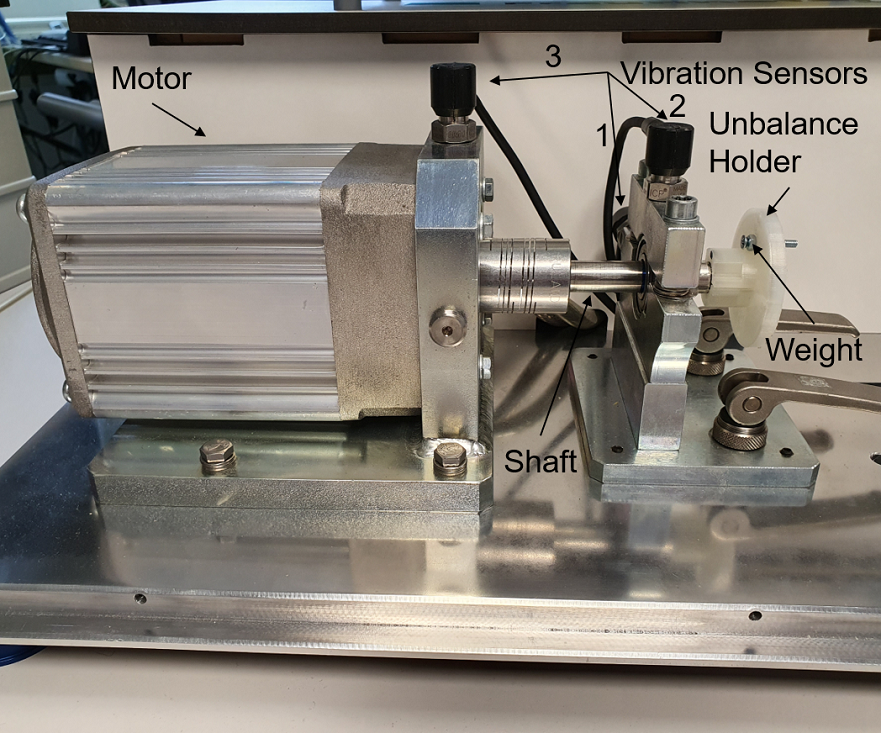}}
\caption{Measurement setup}
\label{setup_image}
\end{figure}
\begin{figure}
\centerline{\includegraphics{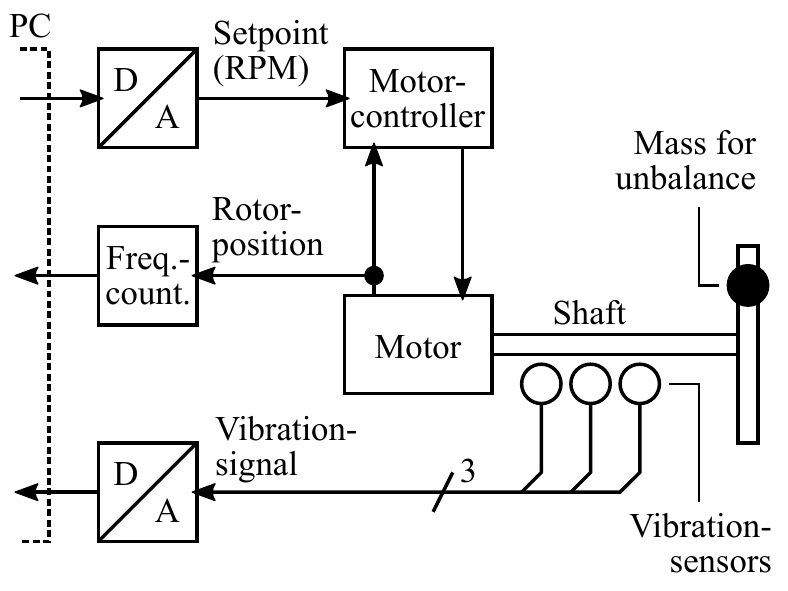}}
\caption{Block diagram of the measurement setup}
\label{setup_block_diag}
\end{figure}

The setup for the simulation of defined unbalances and the measurement of the resulting vibrations is powered by an electronically commutated DC motor (WEG GmbH, type UE 511 T), which is controlled by a motor controller (WEG GmbH, type W2300) and is fixed to the aluminum base plate by means of a galvanized steel bracket. The motor controller allows for a rotation speed between approximately 300 and 2300 revolutions per minute (RPM), which can be continuously adjusted by varying a voltage that is applied to the motor controller. The motor powers a shaft with a diameter of 12 mm which is connected to another shaft of the same diameter and a length of 75 mm by a coupling (Orbit Antriebstechnik GmbH, type PCMR29-12-12-A). This shaft in turn passes through a roller bearing which is clamped in a roller bearing block (material: galvanized steel). The unbalance holder is attached directly behind it. This part was made using a 3D printer (Ultimaker 3, material: nylon) and consists of a disc (diameter: 52 mm) with axially symmetric recesses, in which weights can be inserted to simulate unbalances. Vibration sensors (PCB Synotech GmbH, type PCB-M607A11 / M001AC) are attached to both the bearing block and the motor mounting and are read out using a 4-channel data acquisition system (PCB Synotech GmbH, type FRE-DT9837). As shown in Figure~\ref{setup_block_diag}, the rotation speed of the motor is acquired using a frequency counter in the DT9837, which digitizes the periodicity of the rotor position signal from the motor. A photo of the measurement setup is shown in Figure~\ref{setup_image}.

\section{The Dataset}
\label{dataset_section}

\begin{figure}
\centerline{\includegraphics{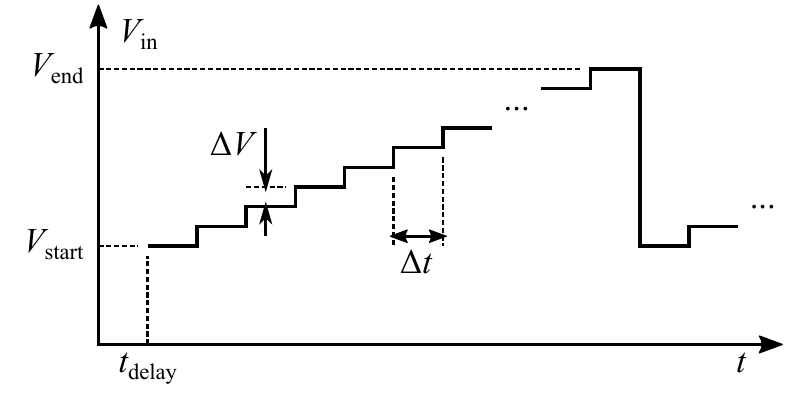}}
\caption{Setpoint for the rotation speed during data acquisition. The setpoint is encoded as a Voltage $V_\textrm{in}$, which is varied according to the diagram above.}
\label{Vin_as_RPM_setpoint}
\end{figure}

\begin{table*}[htb]
\caption{Parameters of used datasets}
\centerline{%
\begin{tabular}{ |l|l|l|l|l|l| }
\hline
ID & Radius & Mass & Unbalance Factor & \multicolumn{2}{l|}{Number of Samples} \\
& [\si{\milli\metre}] & [\si{\gram}] & [\si{\milli\metre\gram}] & Development & Evaluation \\ \hline
0D / 0E&-& 0 & 0 &6438&1670\\
1D / 1E&$14\pm0.1$ &$3.281\pm 0.003$&$\num{45.9 \pm 1.4}$&6434&1673\\
2D / 2E&$18.5\pm0.1$ &$3.281\pm 0.003$&$\num{60.7 \pm 1.9}$&6434&1669\\
3D / 3E&$23\pm0.1$ &$3.281\pm 0.003$&$\num{75.5 \pm 2.3}$&6430&1672\\
4D / 4E&$23\pm0.1$ &$6.614\pm 0.007$&$\num{152.1 \pm 2.3}$&6430&1675\\
\hline
\end{tabular}%
} % end of centerline
\label{measurements_table}
\end{table*}
\begin{figure*}[htbp]
\centerline{\includegraphics[width=1.0\linewidth]{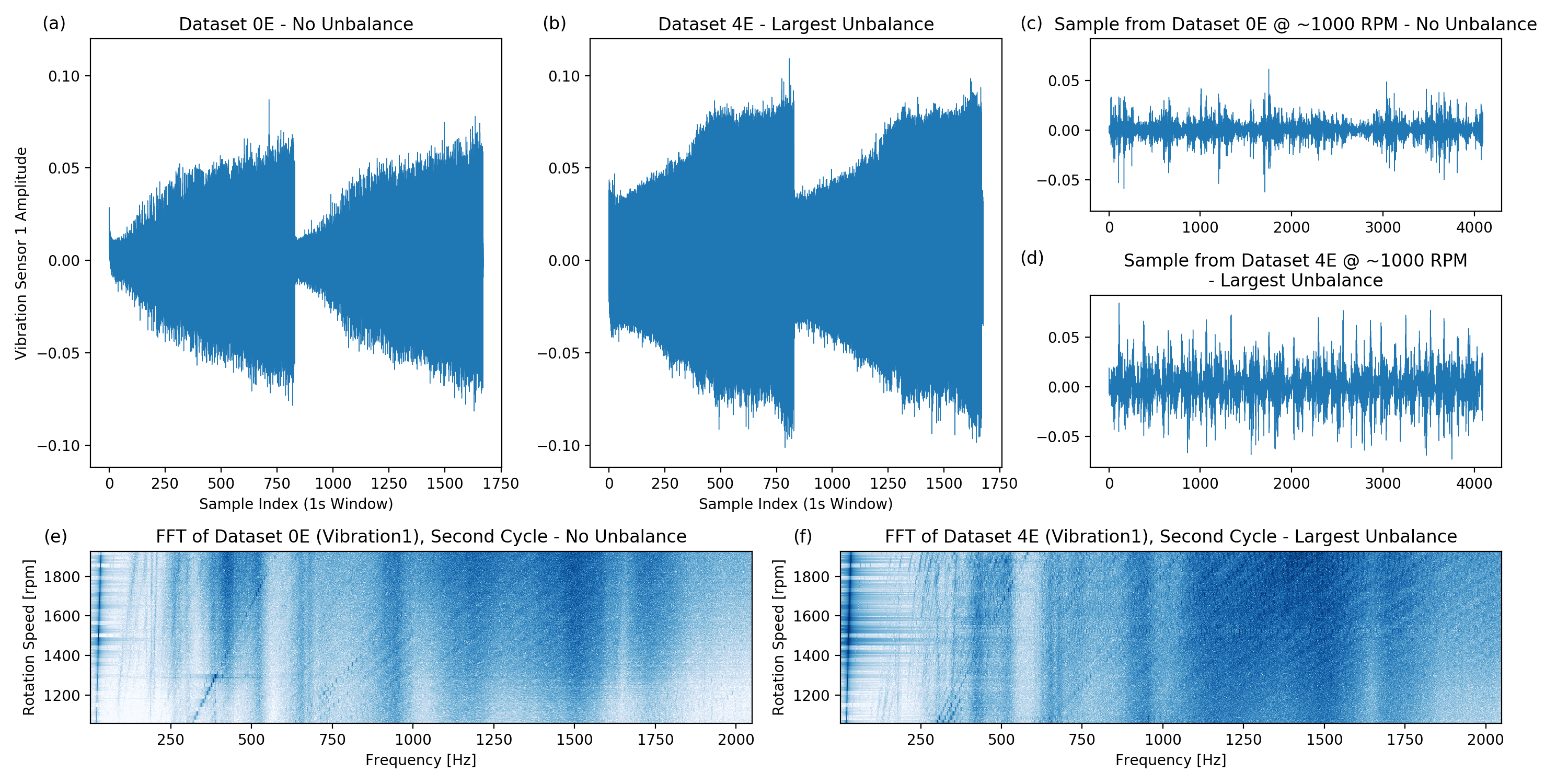}}
\caption{Example measurements from the dataset: Data from vibration sensor 1 for a complete measurement for the case of no unbalance and the largest unbalance ((a) and (b), respectively). For both cases, also a one second sample is extracted ((c) and (d), respectively), as well as the FFT transformation of the second measurement cycle ((e) and (f), respectively).}
\label{dataset_overview}
\end{figure*}

Using the setup described in Section \ref{measurement_section}, vibration data for unbalances of different sizes was recorded. By varying the level of unbalance, different levels of difficulty can be achieved, since smaller unbalances obviously influence the signals at the vibration sensors to a lesser extent. Several further requirements were taken into account: The dataset should be reproducible, relevant for industrial applications and it should represent a use case that is as realistic as possible. This requires the recording of vibration data for varying rotational speed, as an unbalance detector might have to work under varying conditions in some industrial applications. To ensure a high level of (re-)usability of the dataset, we provide tabular data in the \textit{csv}-format.

In total, datasets for 4 different unbalance strengths were recorded as well as one dataset with the unbalance holder without additional weight (i.e. without unbalance). Each dataset is provided as a \textit{csv}-file with five columns:
\begin{description}[\IEEEsetlabelwidth{Measured\_RPM}]
\item[V\_in] The input voltage to the motor controller $V_\textrm{in}$ (in $\textrm{V}$),
\item[Measured\_RPM] the rotation speed of the motor (in $\textrm{RPM}$; computed from speed measurements using the DT9837),
\item[Vibration\_1] the signal from the first vibration sensor,
\item[Vibration\_2] the signal from the second vibration sensor, and
\item[Vibration\_3] the signal from the third vibration sensor.
\end{description}
The sampling rate in each column amounts to 4096 values per second (the rotation speed has been upsampled accordingly).

In order to enable a comparable division into a development dataset and an evaluation dataset, separate measurements were taken for each unbalance strength, respectively. This separation can be recognized in the names of the \textit{csv}-files, which are of the form ``\verb,1D.csv,'': The digit describes the unbalance strength (``\verb,0,'' = no unbalance, ``\verb,4,'' = strong unbalance), and the letter describes the intended use of the dataset (``\verb,D,'' = development or training, ``\verb,E,'' = evaluation).

The unbalance on the measurement setup was completely dismantled and reassembled between the measurement of the development and the evaluation datasets. This causes an additional variability between them and increases the significance of the evaluation of the algorithms that are to be trained on the data. For the development datasets, the motor voltage $V_\textrm{in}$ was increased from $V_\textrm{start} = 2.0\,\textrm{V}$ to  $V_\textrm{end} = 10.05\,\textrm{V}$ in steps of $\Delta V = 0.05\,\textrm{V}$; see Fig.~\ref{Vin_as_RPM_setpoint}. For the evaluation datasets, the motor voltage was increased in steps of $\Delta V = 0.1\,\textrm{V}$ from $V_\textrm{start} = 4.0\,\textrm{V}$ to $V_\textrm{end} = 8.1\,\textrm{V}$. At each step the motor voltage value is kept constant for $\Delta t = \SI{20}{\second}$. The voltage profiles were run through twice for each data record. The rotation speed of the motor was found to be approximately
\begin{equation*}
\frac{n}{\textrm{RPM}} \approx 212 \cdot \frac{V_\textrm{in}}{\textrm{V}} + 209
\end{equation*}
within the range $2\,\textrm{V} \le V_\textrm{in} \le 10\,\textrm{V}$.

An overview of the parameters of the recorded datasets can be found in Table \ref{measurements_table}. This includes the masses and radii for all the used unbalances. Since the absolute value of the centrifugal force $\vec{F}_\textrm{Cf}$ as a function of the rotation speed $\omega$ can under a point mass approximation be expressed as \[
\left|\vec{F}_\textrm{Cf}(\omega)\right| = mr\omega^2, \] the product of the mass $m$ and the radius $r$ is a direct measure of the unbalance strength. In Table \ref{measurements_table} there is also a column called \textit{Number of Samples} for the development and the evaluation dataset. To calculate these values, the first 50,000 values from each dataset were removed (approx. 10 s) and the rest was then divided into windows, each corresponding to one second or 4096 values. The number of samples, therefore, equals the time range of the usable data of each measurement. When calculating prediction accuracies based on these samples, though, it has be taken into account that they are part of a continuous measurement and are therefore not completely independent from each other. Nevertheless, these accuracies are useful to compare the classification performance of different algorithms on the given datasets.

An overview of the amplitude progression at vibration sensor~1 over an entire measurement in the cases of no unbalance and the largest unbalance is shown in Figures~\ref{dataset_overview}(a) and \ref{dataset_overview}(b). For each of the two cases an example curve for one second as well as the FFT transformation of a part of the measurement data is also depicted (Figure~\ref{dataset_overview}(c)-(f)).

\section{Classification of the Unbalance State}
\subsection{Approach 1: Convolutional Neural Network on Raw Sensor Data}
\label{conv_section}
\begin{figure}[htbp]
\centerline{\includegraphics[width=0.5\linewidth]{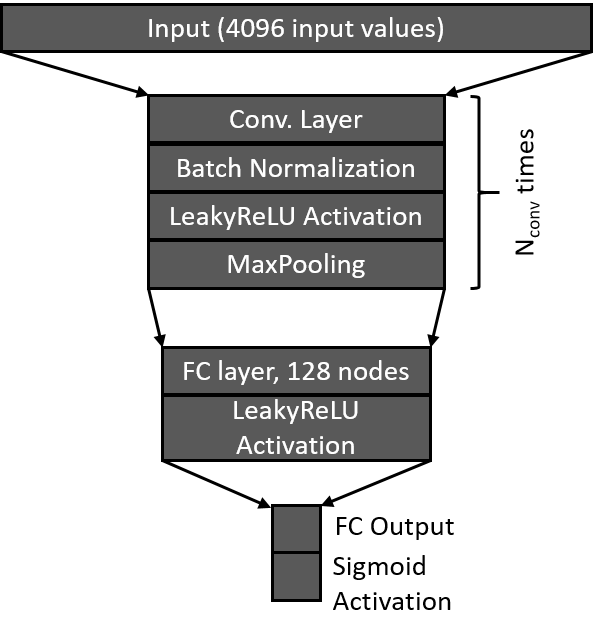}}
\caption{Sketch of the used neural network architecture for the classification of the raw vibration samples. N\textsubscript{conv} describes the number of hidden convolutional and pooling layers used.}
\label{conv_architecture}
\end{figure}

Convolutional Neural Networks (CNNs) are able to recognize patterns in data and to perform classification tasks based on these recognized patterns. An unbalance classification with CNNs, which receive the windowed data directly as input, is therefore promising. The advantage here is that no further data preprocessing is necessary and the effort involved in creating the algorithm is therefore comparatively low. Windowed samples from the data stream `Vibration\_1' were directly used as input. Figure \ref{conv_architecture} shows the CNN architecture used for the classification in this study. 
\begin{figure*}[htbp]
\centerline{\includegraphics[width=1.0\linewidth]{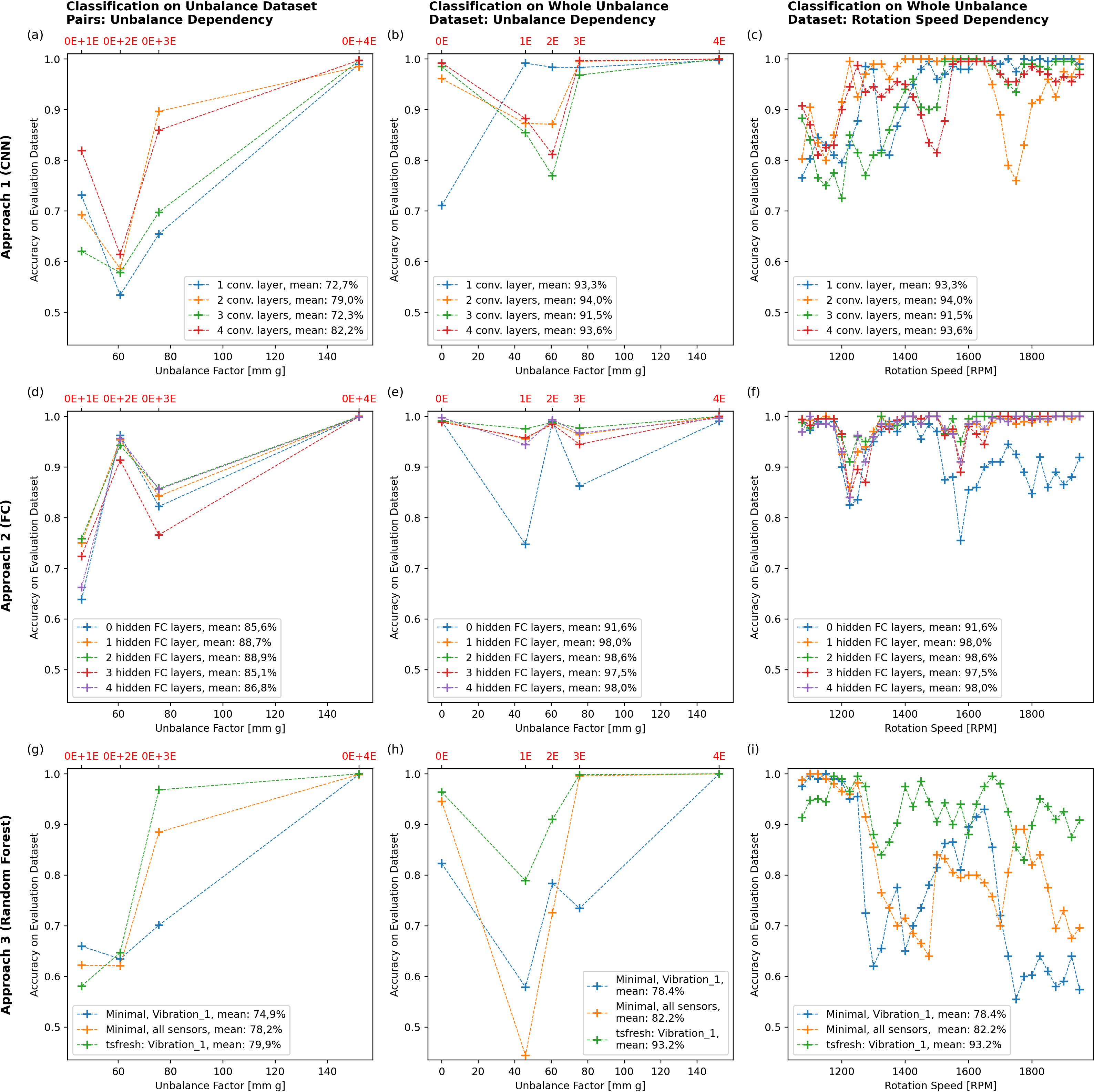}}
\caption{Evaluation accuracies of the used classification approaches 1 (a-c), 2 (d-f) and 3 (g-i) on the task whether or not an unbalance exists. The results for models trained and tested using pairs of the dataset without unbalance and one dataset of a single unbalance strength are plotted in the first column (a,d,g). The results for models trained and tested using all measured unbalance strengths are shown in the second column (b,e,h). The models from the second column are additionally evaluated as a function of the rotation speed (third column, (c,f,i)). In the first two columns, the corresponding IDs of the datasets that were used for the evaluation are marked in red above the diagrams. Connections between the data points are only shown for the sake of clarity and do not imply the continuity of the observed relationships.}
\label{results_all}
\end{figure*}
Since it was shown that an over-parameterization of a neural network with regard to the number of training samples can have a positive effect on the overall performance \cite{belkin2019}, the depth of the network and thus the number of model parameters was varied. This was achieved by varying the number of convolutional blocks $N_{conv}$, consisting of a convolutional layer, batch normalization, activation function and max pooling. After the convolution blocks, one fully-connected (FC) layer leads to the final output layer. Since only the task, whether or not an unbalance is present was used for classification, the output layer consists of one single node with a sigmoid activation.

To better monitor the training process, the development dataset was randomly divided into $\SI{90}{\percent}$ training data and $\SI{10}{\percent}$ test data. During the training phase, the error function on the training data was minimized. The error function based on the test data was also monitored during the training. The model with the lowest error on the test data was kept in each case and afterwards tested on the corresponding evaluation datasets of the same unbalance factors.

For a first attempt, the classification of whether an unbalance is present or not was trained using the data record without unbalance and only one single data record with unbalance each time. Afterwards, all trained models were tested on the corresponding evaluation datasets of the same unbalance factors. The resulting accuracies are shown in Figure~\ref{results_all}(a).

Overall, a rather weak prediction accuracy can be observed in this classification task. It is striking that in particular the second smallest unbalance can hardly be distinguished from the not unbalanced case (dataset pair `0E' and `2E'). With a high prediction accuracy this is only possible for the largest unbalance (dataset pair `0E' and `4E'). With regard to the depth of the CNN, the best results are achieved with 4 and 2 convolutional blocks, with an average of $\SI{79.0}{\percent}$ and $\SI{82.2}{\percent}$ respectively.

In a further experiment, not only one unbalance strength at a time was used as training data, but all. Classifications were nevertheless made as to whether there was an unbalance or not. Therefore there is also one accuracy score describing the performance of the classification algorithm on this task. However, to gain insights into the distribution of correct and incorrect classifications, the resulting models were additionally evaluated in relation to the individual data records, resulting in one accuracy score per unbalance class (plotted in Figure~\ref{results_all}(b)). With this classification task, a significantly better performance of the algorithms used can be observed. On the one hand, this is obviously due to the larger amount of training data available for each individual training. On the other hand, this variant also trains higher variability in the exact mounting of the unbalance. Since the unbalances have been completely disassembled between the measurement of the development and the evaluation datasets, minor changes in the vibration behavior of the entire system can be caused. As in the previous experiment, the best results are achieved with CNNs of 2 ($\SI{94.0}{\percent}$) and 4 ($\SI{93.6}{\percent}$) convolution blocks. With only one convolution block a deviating behavior is obtained: A high detection accuracy, even with small unbalances, is achieved by a reduced detection accuracy of the unbalanced case.

The same trained models were also evaluated as a function of the rotation speed (shown in Figure \ref{results_all}(c)). It can be seen that there are areas, for example around 1600 RPM, where all algorithms have a high prediction accuracy and areas in which this accuracy is low for all models (for example around 1100 RPM). In other areas, however, the prediction accuracy is widely spread (around 1500 RPM).
\subsection{Approach 2: Fully-Connected Neural Network on FFT-transformed Data}
\begin{figure}[htbp]
\centerline{\includegraphics[width=0.5\linewidth]{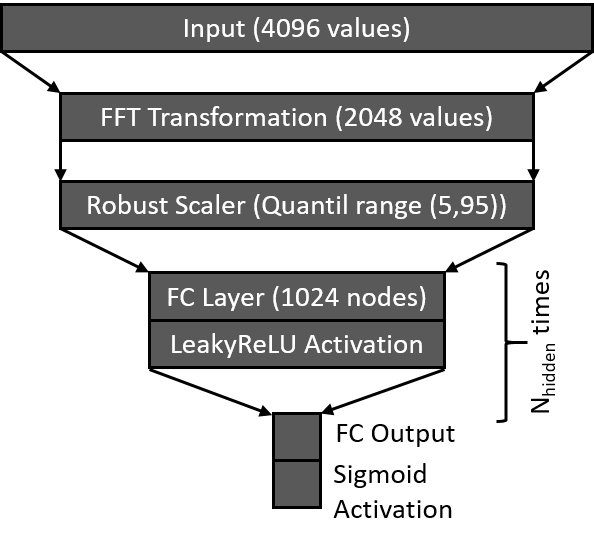}}
\caption{Sketch of the used neural network architecture for the classification of the FFT-transformed vibration samples. N\textsubscript{hidden} describes the number of hidden FC layers used.}
\label{fc_architecture}
\end{figure}

For this approach, the FFT was calculated for each of the windows of one second or 4096 values of the first vibration sensor stream (`Vibration\_1'). According to the Shannon-Nyquist sampling theorem, this results in 2048 physically meaningful Fourier coefficients for each window, which can be used for classification. Again, development dataset transformed in this way was randomly divided into $\SI{90}{\percent}$ training data and $\SI{10}{\percent}$ test data. Afterwards, the FFT data were scaled as follows: For each Fourier coefficient, the respective median and interquantile spacing of quantiles 5 and 95 was calculated based on the extent of the training dataset (2048 values for the median and the interquantile spacing, respectively). The median values were then subtracted from the FFT values and the result was divided by the interquantile values. Fully connected (FC) neural networks were then trained on the training data. An illustration of the used network architectures is shown in Figure~\ref{fc_architecture}. The input consisting of 2048 Fourier coefficients in each sample was followed by N\textsubscript{hidden} hidden and fully connected layers with LeakyReLU activation and the output layer. Neural networks of this type with zero (equivalent to logistic regression) to four hidden layers were trained using the respective training data. As described in approach 1 (Section \ref{conv_section}), in the first experiment the dataset of the unbalance-free case and the datasets with unbalance were paired and the respective models were trained  based on these datasets and evaluated on the corresponding dataset pairs from the evaluation data. 

The resulting accuracies are shown in Figure~\ref{results_all}(d). It is apparent that the trained models are only partially able to accurately conduct the classification between zero and the smaller unbalances. While the largest unbalance could be classified correctly in almost every case, the picture is inconsistent in the remaining cases. The data record `2E' can also be classified with a high accuracy, while this in turn works worse for the data record `3E'. A monotonically increasing prediction accuracy as a function of the unbalance strength was expected. One reason for the deviation observed here could be again effects caused by reassembling of the unbalance after each measurement. Additionally, there is a slight trend visible, that neural networks with one or two layers reach a better overall performance in this task, possibly a better generalization could be achieved in these cases.

In the next experiment, again all datasets instead of pairs of datasets were used and classification was conducted, whether or not there was an unbalance at all. The results are depicted in Figure~\ref{results_all}(e). The number of hidden FC layers was again varied between 0 and 4. Overall, an accuracy of 0.916 (zero hidden layers) to 0.986 (two hidden layers) was achieved on the evaluation dataset for the classification task. While the largest unbalance is recognized almost perfectly by all methods, there is no clear tendency for the other unbalance strengths, similar to the previous experiment. 
As with the CNN approach, the larger amount of training data also leads to better performance overall.

For the rotation speed dependent evaluation (Figure~\ref{results_all}(f)), it can be seen that all models have a drop in prediction accuracy in the ranges around 1200 RPM and 1550 RPM. Outside these ranges, all algorithms except those with zero hidden layers achieve an accuracy of almost $\SI{100}{\percent}$. One reason for the worse performance in the described ranges could be resonant oscillations of the measurement setup, resulting in a reduction of the signal-to-noise ratio of the signals caused by the unbalance.

\subsection{Approach 3: Random Forest on Automatically Extracted Timeseries Features}
In order to compare the generalization ability of all employed algorithms to a common baseline, and to find out to what extent a higher computational effort has an impact on a possibly improved prediction accuracy, classification was carried out using a minimum set of features. This small feature set consists of the mean of the `Measured\_RPM' values, as well as the standard deviation and the kurtosis of the vibration values, which were calculated for each of the previously partitioned windows of the datasets. This feature calculation was carried out in two variants: First, standard deviation and kurtosis were only calculated for `Vibration\_1', resulting in a total of 3 features (including the mean of the `Measured\_RPM' values). In the second variant, all three vibration sensors were used (7 features in total). Both variants shall be denoted to as `minimal features'. A Random Forest model was then trained on these minimal features. As with the previous classification approaches, the classification training was conducted once with dataset pairs consisting of one unbalance strength and the unbalance-free case each and once with all existing unbalance strengths. The classification results of the evaluation are shown in the Figures \ref{results_all}(g)-(i). It can be seen that the highest unbalance can be detected almost perfectly by using only 3 features in both experiments. Using 7 features and trained on the whole dataset, even the dataset `3E' can  be classified close to \SI{100}{\percent} accuracy. With the smaller unbalances, on the other hand, there is a significant decrease in the prediction accuracy and also the unbalance-free case can only be detected to \SI{82.2}{\percent} (3 features) or \SI{94.6}{\percent} (7 features) in the second experiment (Figure \ref{results_all}(h)). When looking at the rotation-speed-dependent evaluation, the high accuracy below 1200 RPM is particularly striking. In this range, the classification with the minimum set of features even achieves a significantly better result than with approach 1.

Besides the mentioned minimal features, the Python package \textit{tsfresh} offers the possibility of computing a much wider range of features describing time series \cite{christ2018}. Using \textit{tsfresh} (version 0.14.1), 748 features belonging to the class \texttt{EfficientFCParameters()} were extracted for `Vibration\_1' and afterwards used as input for a random forest algorithm. Since the classification task and algorithm remained the same, and only the number of input features changed, the prediction results on the evaluation dataset are depicted in the Figures \ref{results_all}(g)-(i), as well (green curve). In particular, a significant improvement in the detection rate for the smaller unbalances compared to the minimal features causes that, with a total prediction accuracy of \SI{93.2}{\percent} when trained on all unbalance strengths and a mean prediction accuracy of \SI{79.9}{\percent} when trained with the dataset pairs, a level similar to that of the CNNs in approach 1 (Section \ref{conv_section}) is achieved overall.

\subsection{Approach 4: Hidden Markov Model}
\begin{figure}
\centerline{\includegraphics{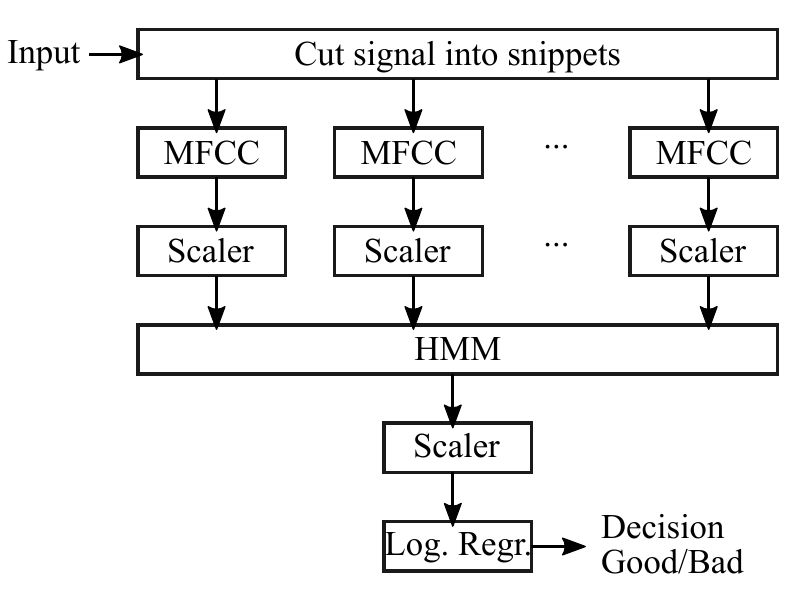}}
\caption{Block diagram of the unbalance detector using HMM and MFCCs.}
\label{HMM_Block_Diag}
\end{figure}

One can obtain an outline of hidden markov models (HMMs) from the article \cite{Rabiner86}. Beyond speech recognition (e.g. in the \emph{Sphinx} system \cite{Lee90,Walker04}), HMMs have also been used in biology (e.g. protein structure and genome research \cite{Eddy96}), sports (e.g. recognition of sports activities \cite{Hausberger16}), and other use cases. In the field of condition monitoring, HMMs have e.g. been employed for the detection of defective roller bearings \cite{Marwala07}. The latter paper has had a certain influence on the design of the unbalance detector, that is used in the present section:

A possible approach to determine the unbalance state using a hidden markov model (HMM) is shown in Figure~\ref{HMM_Block_Diag}: The input signal (4096 consecutive values from `Vibration\_1' were used) is cut into (possibly overlapping) snippets of a fixed length. For each snippet, the mel-frequency cepstral components (MFCCs) are computed as features, which are then input into a HMM that is trained to recognize data without unbalance. To facilitate the interpretation of the HMM output, logistic regression is used to decide whether a given input signal results from a measurement with or without unbalance. The scalers in Figure~\ref{HMM_Block_Diag} simplify the training process.

\begin{figure*}[ht]
\begin{center}
\includegraphics[width=\textwidth]{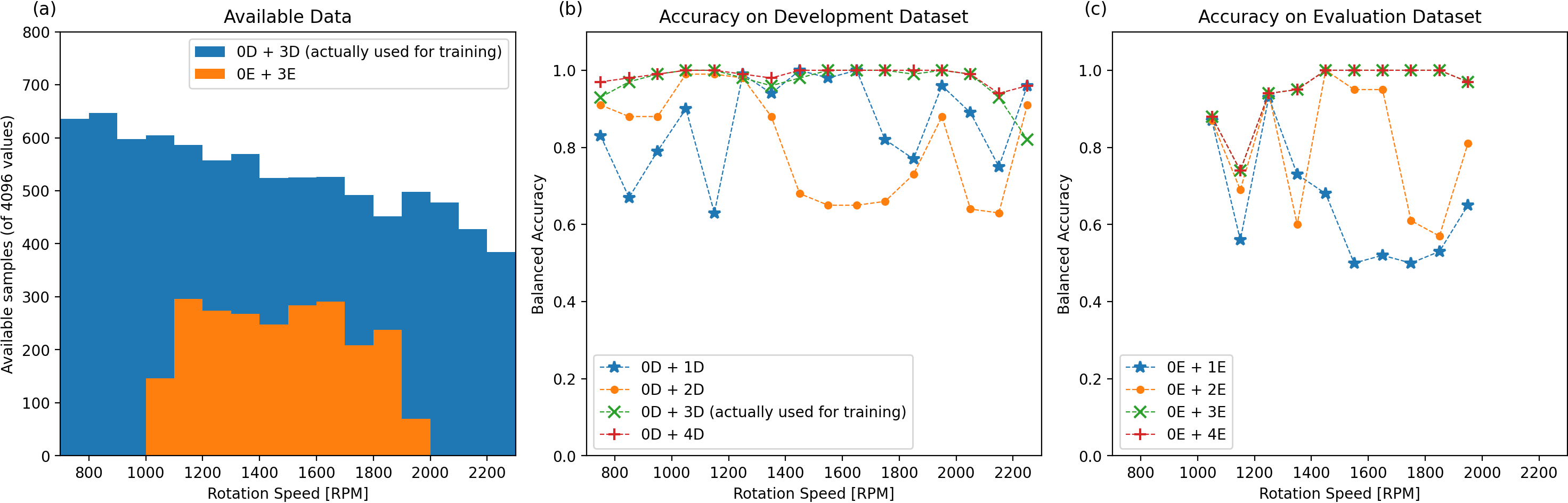}
\end{center}
\caption{Classification results using the Hidden Markov Model with MFCC features}
\label{HMM_Results}
\end{figure*}
Because the MFCC features are sensitive to variations in the rotation speed, it was decided to train several models for different speeds. The training data was therefore assembled in the following way: One-second samples of the `Vibration\_1' signal (from the `0D' and `3D' datasets) were selected such that the speed (`Measured\_RPM') is always within a certain interval. The training data is then randomly split into three sets. One set is used to train the first scaler and the HMM (using data from `0D' only). The second set is used to train the second scaler and the logistic regression to recognize measurements with unbalance. The third set is used to determine hyperparameters (number of MFCC features, number of HMM states, snippet length and overlap), that maximize the balanced accuracy.

During development of the HMM approach, it was noticed that the MFCC features within one-second samples of the measurement without unbalance appear to be relatively stationary. And the hyperparameter optimization often results in the use of only one (!) HMM state. The problem at hand might therefore be inadequate for HMMs (which are usually applied to instationary processes).

Figure~\ref{HMM_Results} shows the results of the HMM approach to unbalance detection. It can be seen, that the balanced accuracy is in the range
\begin{itemize}
% Note: Identical results for 3E and 4E can happen; e.g. if all "bad" samples are classified correctly!
\item 0.56--0.93, mean = 0.65 (for the union of `0E' and `1E'),
\item 0.57--1.00, mean = 0.80 (for the union of `0E' and `2E'),
\item 0.74--1.00, mean = 0.95 (for the union of `0E' and `3E'), and
\item 0.74--1.00, mean = 0.95 (for the union of `0E' and `4E').
\end{itemize}

\section{Summary and Outlook}
For this study, a dataset with vibration data for the classification of unbalance on a rotating shaft with variable speed and unbalance strength was created. Various approaches to solve the associated classification task were tested. The largest unbalance could be detected by all algorithms with almost perfect prediction accuracy, even if only 3 characteristic values per sample were used for the classification. With the smaller unbalances, on the other hand, wider variations between the different approaches were found. The best way to classify the dataset was to use an FC network with two hidden layers, which received the scaled FFT-transformed vibration data as input. Measured on the entire evaluation dataset, \SI{98.6}{\percent} of the cases could be classified correctly.
In addition, the examined models showed a very different behavior regarding the dependence on the speed. In future studies, this behavior could be exploited by building ensembles of different models to further increase the prediction accuracy. Strengths and weaknesses of individual models in the different speed ranges would then at least partially compensate each other. Moreover, for the further improvement of the models as well as the understanding of the classifications, for example in a productive company, efforts with regard to enabling a traceability and explainability of the models used are necessary.

\end{document}